\documentstyle[12pt,titlepage]{article}

\font\grande=cmr10 scaled \magstep4
\font\medio=cmr10 scaled \magstep2
\outer\def\beginsection#1\par{\medbreak\bigskip
      \message{#1}\leftline{\bf#1}\nobreak\medskip
\vskip-\parskip
      \noindent}

\def\laq{\raise 0.4ex\hbox{$<$}\kern -0.8em\lower 0.62
ex\hbox{$\sim$}}
\def\gaq{\raise 0.4ex\hbox{$>$}\kern -0.7em\lower 0.62
ex\hbox{$\sim$}}

\begin{document}
\bibliographystyle {unsrt}
\newcommand{\pa}{\partial}
\newcommand{\rhob}{{\bar \rho}}
\newcommand{\prb}{{\bar p}}

\titlepage
\begin{flushright}
CERN-TH/95-85 \\
\end{flushright}
\vspace{15mm}
\begin{center}
{\grande Primordial Magnetic Fields}\\
\vspace{5mm}
{\grande from String Cosmology}

\vspace{10mm}

M. Gasperini \\
{\em Dipartimento di Fisica Teorica, Via P. Giuria 1, 10125 Turin,
Italy} \\
M. Giovannini and G. Veneziano \\
{\em Theory Division, CERN, CH-1211 Geneva 23, Switzerland} \\
\end{center}
\vspace{10mm}
\centerline{\medio  Abstract}

\noindent
Sufficiently large seeds for generating the observed
(inter)galactic magnetic fields emerge naturally in string
cosmology from the amplification of electromagnetic
vacuum fluctuations  due  to a dynamical dilaton background.
The success of the mechanism
depends crucially on two features of the so-called
pre-big-bang scenario, an early epoch
of dilaton-driven inflation at very small coupling, and  a
sufficiently long intermediate stringy era preceding
the standard radiation-dominated evolution.

\vspace{5mm}
\vfill
\begin{flushleft}
CERN-TH/95-85 \\
April 1995 \end{flushleft}

\newpage

\setcounter{equation}{0}

It is widely believed that the observed galactic (and intergalactic)
magnetic fields, of microgauss  strength, are generated and
maintained by
the action of a cosmic dynamo \cite{Parker}. Any dynamo model,
however,
 requires a primordial seed field; in spite of many
attempts \cite{Turner}-\cite{Harrison},
it is fair to say that no compelling mechanism able to
generate the required seed
field (coherent over the Mpc scale, and with an
energy density to radiation density ratio
$\rho_{B}/\rho_{\gamma} \gaq 10^{-34}$) has yet been
suggested.

A priori, an appealing mechanism for the origin
 of the seed field is the cosmological amplification of the
 vacuum quantum fluctuations
of the electromagnetic field, the same kind of mechanism
which is believed to generate primordial metric and energy
density perturbations \cite{Grishchuk}. The minimal coupling of
photons
to the  metric background is, however, conformally invariant
(in $d=3$ spatial dimensions). As a consequence, a
cosmological evolution involving a conformally flat metric
(as it is effectively the case in  inflation)
cannot amplify magnetic fluctuations, unless conformal
 invariance is broken.
Possible attempts to generate large enough seeds thus include
considering exotic higher-dimensional scenarios, or  coupling
non-minimally the electromagnetic field to the
background curvature
\cite{Turner} with some ``ad hoc" prescription, or breaking
conformal invariance at the quantum level through the so-called
trace anomaly \cite{Dolgov}.

In critical superstring theory the
electromagnetic field $F_{\mu\nu}$ is coupled not only
to the metric ($g_{\mu\nu}$), but also to the dilaton background
($\phi$).
In the low energy limit such interaction is represented by the
string effective action \cite{Lovelace}, which reads, after
reduction from ten to four external dimensions,
\begin{equation}
S=- \int d^4x\sqrt{-g}e^{-\phi}( R +
\partial_{\mu} \phi \partial^{\mu} \phi
+ \frac{1}{4} F_{\mu\nu}F^{\mu\nu})
\label{action4}
\end{equation}
were  $\phi = \Phi - \ln{V_6} \equiv \ln (g^2)$ controls the
tree-level four-dimensional gauge coupling ($\Phi$ being the
ten-dimensional dilaton field, and $V_6$ the volume of the
six-dimensional compact internal space).

In the inflationary models based on the above
effective action \cite{Veneziano,Gasperini}
the dilaton background is not at all constant, but
undergoes an accelerated evolution
from the string perturbative vacuum ($\phi= -\infty$) towards the
strong
coupling regime, where it is expected to remain frozen at its
present value. In this
context, the quantum fluctuations of
the electromagnetic field can
thus
be amplified $\it{directly}$ through their coupling to the dilaton,
according
to eq.(\ref{action4}). In the following we will discuss the
conditions
under which such mechanism is able to produce
large enough primordial magnetic fields to seed the galactic
dynamo (a scalar-vector coupling similar to that of
eq.(\ref{action4}) was previously discussed in \cite{Ratra}, but
$\phi$ was there identified with the conventional inflaton
undergoing a dynamical evolution much different from the
dilaton evolution considered here).

Let us first define a few important parameters of the inflationary
scenario (also called "pre-big-bang" scenario)
discussed in \cite{Gasperini}. The phase
of growing  curvature and dilaton coupling
($\dot H>0$, $\dot\phi>0$), driven by the kinetic energy of the
dilaton field,  is correctly described in terms of
 the lowest order string effective action only
up to the conformal time $\eta=\eta_{s}$  at which the curvature
reaches the string scale $H_{s}=\lambda_{s}^{-1}$ ($
\lambda_{s}\equiv
\sqrt{\alpha^{\prime}}$ is the fundamental length of string theory).
A first important parameter of this cosmological model is thus the
value $\phi_s$ attained by the dilaton at $\eta=\eta_{s}$.
Provided such
 value is sufficiently negative (i.e. provided the coupling
$g=e^{\phi/2}$ is sufficiently small to be still in the
perturbative region at $\eta=\eta_{s}$), it is also arbitrary, since
there
is no perturbative potential to break invariance under shifts of
$\phi$.
 For $\eta >\eta_{s}$ high-derivatives terms (higher
orders in $\alpha^{\prime}$)
become important in the string effective action,
and the background enters a genuinely ``stringy" phase of
 unknown duration. It was shown in
\cite{Brustein} that it is impossible to have a graceful
exit to  standard cosmology without such an intermediate
stringy phase. An  assumption of string cosmology
is that the stringy phase eventually ends at some conformal time
$\eta_1$
in the strong coupling regime.  At this time the dilaton,
feeling a non-trivial potential, freezes to its present constant
value
$\phi=\phi_{1}$ and the standard radiation-dominated era starts.
The total duration $\eta_1/\eta_s$, or the total red-shift $z_s$
encountered during
 the stringy epoch (i.e. between $\eta_s$ and $\eta_1$),
will be the second crucial parameter besides $\phi_{s}$
entering  our discussion. For the purpose of this paper, two
parameters are enough to specify completely our model of
background, if we accept that during the string phase the
curvature stays controlled by the string scale, that is $H\simeq g
M_p=\lambda_{s}^{-1}$ ($M_p$ is the Planck mass) for
$\eta_s<\eta <\eta_1$.

We will work all the time in the String (also
called Brans-Dicke) frame, in which test strings move along
geodesic surfaces. In this frame the string scale $\lambda_{s}$
 is constant, while the Planck scale
$\lambda_{p}= e^{\frac{\phi}{2}} \lambda_{s}$ grows from zero
(at the initial vacuum) to its present value,
 reached at the end of the
string phase. We have explicitly checked that all our results
also follow in the more commonly used (but less natural
in a string context) Einstein frame, in which the gravi-dilaton
action is diagonalized in standard canonical form. Moreover, our
parameterization in terms of $\phi_s$ and $\eta_1/\eta_s$ is
completely frame-independent, since the dilaton field and the
conformal time coordinate are the same in the String and
Einstein frame \cite{Gasperini}.

We shall now consider, in the above background, the amplification
of the quantum fluctuations of the electromagnetic field,
assuming that, at the very
beginning, it was in its vacuum state. In a four-dimensional,
conformally flat background,
the  Fourier modes $A_{k}^{\mu}$ of the (correctly
normalized) variable
 corresponding to the standard
electromagnetic field, and obeying canonical commutation
relations,
 satisfy the equation
\begin{equation}
A_{k}^{\prime\prime}+
[k^2-V(\eta)]A_{k}=0~~~~,~~~~V(\eta)= g(g^{-1})^{\prime\prime}~~~~,~~~~
g(\eta)\equiv e^{\frac{\phi}{2}}
\label{equazione}
\end{equation}
This equation is valid for each polarization component, and
 is obtained from the action (\ref{action4}) with the gauge
condition
$\partial_{\nu}[e^{-\phi}\partial^{\mu}(e^{\frac{\phi}{2}}
A^{\nu})]=0$
(a prime denotes
differentiation with respect to $\eta$).
Note the analogy with the equation for the tensor
part of metric perturbations \cite{Grishchuk}. The latter have the
same form
 as (\ref{equazione})
with the inverse of the string coupling, $g^{-1}$, replaced simply
by the Einstein-frame scale factor $a_E = a e^{-\phi/2}$,
expressed in conformal time.  Equation (\ref{equazione})
clearly shows the absence of parametric
amplification whenever $\phi$ is constant.

The effective potential $V(\eta)$ grows from zero like
$\eta^{-2}$, for $\eta \rightarrow 0_-$, in the phase of
dilaton-driven inflation (where \cite{Veneziano,Gasperini} $\phi
\sim -\sqrt 3 \ln|\eta|$), is expected to reach some maximum
value during the string phase, and then goes rapidly to zero at
the beginning of the radiation-dominated era (where $\phi =$
const).
 The approximate solution of eq.
(\ref{equazione}), for a mode $k$ "hitting" the effective
potential barrier at $\eta=\eta_{ex}$, and
with initial conditions corresponding to vacuum
fluctuations, is given by:
\begin{eqnarray}
 A_{k} &=&{e^{-ik\eta}\over \sqrt{k}}\;\;~~~,~~~~~
\,\,\,\,\,\,\,\,\,\,\,\,\,\,\,\,\,\,\,\,\,\,\,\,\,\,\,\,\,\,\,\,\,\,\,\,\,\,\,
\eta < \eta_{ex} \;
 \nonumber \\
A_{k} &=& g^{-1}(\eta)
 [C_k + D_k \int^{\eta} d\eta'~~ g^2(\eta^{\prime})] \; \;
{}~~~~,~~~~~\eta_{ex} < \eta < \eta_{re} \; \nonumber \\
A_{k} &=& {1\over \sqrt{k}}[ c_+(k) e^{-ik\eta} +  c_-(k)
e^{ik\eta}]
\;\;~~~~~,~~~~~~~~ \eta > \eta_{re}
\label{soluzione}
\end{eqnarray}
where $\eta_{ex}$ and $\eta_{re}$ are the times of exit and
reentry of
the comoving scale associated with $k$, defined by the
conditions  $k^2 = |V(\eta_{ex})| = |V(\eta_{re})|$ ( $C, D,
c_{\pm}$ are integrations constants). We are following here the
usual convention for which a mode in the underbarrier region is
referred to, somewhat improperly, as being "outside the
horizon". Hence the names "exit" and "reentry". Moreover, we
are considering a background in which the potential $V(\eta)$
keeps growing in the string phase until the final time $\eta_1$,
so that a mode crossing the horizon during dilaton-driven
inflation remains outside the horizon during the whole string
phase, i.e. $\eta_{re}\geq \eta_1$.

The Bogoliubov coefficients $c_{\pm}(k)$, determining the
parametric amplification of a mode $k<|V(\eta_1)|$, are easily
determined by matching these various solutions. One finds:
\begin{eqnarray}
{2ik}e^{- ik(\eta_{ex} \mp \eta_{re})} c_\pm =
 &\mp &
\frac{g_{ex}}{g_{re}}(-\frac{{g_{re}}^{\prime}}{g_{re}} \mp ik) \pm
\frac{g_{re}}{g_{ex}}(-\frac{{g_{ex}}^ {\prime}}{g_{ex}} +ik) \pm
 \nonumber \\
&\pm &\frac{1}{g_{ex} g_{re}} (-\frac{{g_{ex}}^{\prime}}{g_{ex}}
+ik)  (-\frac{{g_{re}}^{\prime}}{g_{re}} \mp ik)
\int_{\eta_{ex}}^{\eta_{re}} g^2 d\eta
\label{Bog}
\end{eqnarray}
If we remember that reentry occurs during the radiation  epoch
in which the dilaton freezes to a constant value
($g^{\prime}_{re}\simeq 0$), it is easy to estimate  the
complicated  looking expression (\ref{Bog})
and to obtain, for the leading contribution,
the amazingly simple and intuitive result:
\begin{equation}
|c_{-}|\simeq\frac{g_{re}}{g_{ex}}\equiv
e^{-\frac{1}{2}(\phi_{ex}-\phi_{re})}
\label{amplification}
\end{equation}
expressing the fact that the amplification of the electromagnetic
field depends just (to leading order) on the ratio of the gauge
couplings
at reentry and at exit.

The coefficient (\ref{amplification}) defines
the energy density distribution ($\rho_{B}(\omega)$) over
the amplified fluctuation spectrum,
$ d\rho_{B}/d\ln\omega \simeq \omega^4
|c_{-}(\omega)|^2$ , where $\omega= k/a$ is the red-shifted,
present value of the
amplified proper frequency. We are interested in the ratio
\begin{equation}
r(\omega)=\frac{\omega}{\rho_{\gamma}} \frac{d\rho_{B}}
{d\omega}
\simeq
\frac{\omega ^{4}}{\rho_{\gamma}} |c_{-}(\omega)|^2 \simeq
\frac{\omega^{4}}{\rho_{\gamma}}
\left(g_{re}\over g_{ex}\right)^2
\label{r}
\end{equation}
measuring the fraction of electromagnetic energy stored in the
mode
$\omega$ (in particular, for the intergalactic scale,
$\omega_{G}\simeq (1
Mpc)^{-1}\simeq 10^{-14}$Hertz),
 relative to the background radiation energy $\rho_{\gamma}$.

The ratio $r(\omega)$ stays
constant
 during the phase of matter-dominated as well as
radiation-dominated
evolution, in which the universe behaves like a good
electromagnetic
 conductor
\cite{Turner}.
In terms of $r(\omega)$ the condition for a large enough
magnetic field  to seed the galactic dynamo is  \cite{Turner}
\begin{equation}
r(\omega_{G})\gaq 10^{-34}
\label{condition}
\end{equation}
Using the known value of $\rho_{\gamma}$ and $e^{\phi_{re}}$
we thus find, from eqs.(\ref{r}, \ref{condition}):
\begin{equation}
 g_{ex}(\omega_G) < 10^{-33}
\label{r2}
\end{equation}
i.e. a very tiny coupling at the time of exit of the (inter)galactic
scale $\omega_G$ (we may note, incidentally, that it
 looks very unlikely that such a
large ratio of  $g_{re}/g_{ex}$ can be obtained through the trace
anomaly mechanism of Ref. \cite{Dolgov}).

In order see whether or not the previous condition
can be fulfilled we go back to our two-parameter
cosmological model. We recall that, in the string phase, the
curvature stays controlled by the string scale, so that $H\simeq
\lambda_s^{-1}=$ constant in the String frame, and
$a_1/a_s=\eta_s/\eta_1=z_s$.  The
discussion is greatly helped by looking at {\bf Fig.1} where we
plot, on a double-logarithmic scale against the scale factor $a$,
the evolution of the
coupling strength (i.e. of $e^{\phi/2}$) and that of the
``horizon" size (defined here by $a |V|^{-1/2}$, whose behavior
coincides with that of the Hubble radius $H^{-1}$
during the dilaton driven epoch).
The horizon curve has an inverted trapezoidal shape,
corresponding to the fact that
$V=0$ during the radiation era, that $\dot\phi$ and $H$ are
 approximately constant during the string era,
and that, during the dilaton-driven era
\cite{Veneziano,Gasperini},
\begin{equation}
a=(-t)^{\alpha},~~~~~ \alpha = - \frac {1}{\sqrt{3}}
\sqrt{1- \Sigma}, \,\,\,\,\,\,\,\,
  a|V|^{-\frac{1}{2}}\simeq
a(t)\int_{t}^{0} dt^{\prime} a^{-1}(t^{\prime})
\simeq a^{{1\over \alpha}}
\end{equation}
Here
$\Sigma \equiv \sum_{i}{\beta_i^2}$ represents the
possible effect of internal dimensions whose radii $b_i$ shrink
like $(-t)^{\beta_i}$ for $t \rightarrow 0_-$ (for the sake of
definiteness we show in the figure the case $\Sigma=0$).
The shape of the coupling curve corresponds to the fact that the
dilaton
is constant during the radiation era, that $\dot\phi$ is
 approximately constant during the string era, and that it evolves
like
\begin{equation}
g(\eta) = a^{\lambda}, \,\,\,\,\,\,\,\,\,\,
\lambda =
\frac{1}{2}(3+\frac{\sqrt{3}}{\sqrt{1-\Sigma}})
\end{equation}
during the dilaton-driven era \cite{Veneziano,Gasperini}
($\Sigma=0$ is the case shown in the picture).

We can now easily see when a sufficient amplification is achieved.
The galactic scale of length $\omega_G^{-1}$ was about $10^{25}$
in string  (or Planck) units at
the beginning of the radiation era. By definition, at earlier times
it evolves as a straight line with a positive slope on our plot and
thus inevitably
hits the horizon curve sometimes during the string or the
dilaton-driven
era. At that time, the value of $g$ should have been smaller than
$10^{-33}$. One can easily convince her/himself that this is all but
impossible provided:
i) $z_s= a_1/a_s$ is a sufficiently large, and ii) the dilaton
evolution during the
string era is sufficiently fast. For the first condition a red-shift
$z_{s}$ of
$10^{10}$ is necessary, while for the average
 ratio ${\dot \phi / H}$ during the string era a value
below but not too far from the one just before $\eta_s$ is
sufficient.
The combination of i) and ii) also implies that the coupling
 at the onset of the string era has to be smaller
that $10^{-20}$ or so,
which thus supports the scenario advocated in \cite{Brustein} for
the gracious exit problem.

We can  express our results more quantitatively
 by showing the allowed region in the
$g_s - z_s$ plane in order to have  sufficiently large seeds.
Considering the possibility of galactic scale  exit during the string
or the dilaton-driven phase,  we find from eq.(\ref{r}) that
 $r(\omega_G)$ can be expressed, in the two cases, respectively
as:
 \begin{equation}
r(\omega_{G})\simeq \left(\frac{\omega_{G}}
{\omega_{1}}\right)^{4}
e^{-\phi_{ex}(\omega_G)}\simeq
\left(\frac{\omega_{G}}{\omega_{1}}\right)
^{4 +\frac{\phi_{s}}{\ln{z_{s}}}}
{}~~~~~~,~~~\omega_{s}<\omega_{G}<\omega_{1}
\label{romega}
\end{equation}
and
 \begin{equation}
r(\omega_{G})\simeq\left
(\frac{\omega_{G}}{\omega_{1}}\right)^{4-2\gamma}
z_{s}^{-2\gamma} e^{-\phi_{s}}
{}~~~~\,\,\,\,\,\,\,,~~~\omega_{G}<\omega_{s} \label{romega2}
\end{equation}
where $\gamma =\lambda\alpha/(\alpha -1)$,
$\omega_1=H_1a_1/a\simeq 10^{11}$Hertz is the maximal
amplified frequency,
 and
$\omega_{s}=\omega_{1}/z_{s}$. In the previous formulae
(\ref{romega}), (\ref{romega2}) we used the fact that, according
to our model of background, the transition scale $H_{1}$ has to
be of the order of the Planck mass $M_p$, so that
$\rho_{\gamma}(t) \simeq
H_{1}^4[a_1/a(t)]^4  ={\omega_1}^4$.

The resulting limits obtained by imposing eq.(\ref{condition})
are plotted in {\bf Fig.2}, where they provide the right-side
border of the allowed region (the shaded area). According to the
previous spectrum, however, the amplitude of the
electromagnetic perturbations is not constant outside the
horizon, but tends to grows, asymptotically, with a power-like
behavior  (similar effects are known to occur also for metric
perturbations in string cosmology backgrounds \cite{BG}).
 A further
bound on the allowed region thus
emerges in the context of this model, as we must require for
consistency that our perturbative treatment of the vacuum
fluctuations remains valid at all the times,
or, in other words, that the energy density of the
produced photons is smaller than critical also in the radiation era,
i.e. \begin{equation}
 r(\omega) <1 \label{crit1}
\end{equation}
at all $\omega$.  For electromagnetic perturbations crossing
the horizon during the  dilaton-driven period, in
particular, this condition is most stringent at
$\omega=\omega_{s}$.

Using eqs. (\ref{romega}),
(\ref{romega2}) and (\ref{crit1}) we find that the two different
branches of the photon spectrum give the same bounds,  $g_s<1$
and $\log_{10} g_s >- 2\log_{10}z_s$, the latter
determining the left border of the allowed region. Actually, a
slightly tighter bound ($r\laq 0.5$) could also be imposed in order
to avoid perturbations of the standard nucleosynthesis scenario
\cite{Cheng}, but this does not affect in  significant way the
result presented in {\bf Fig.2}, which should be regarded, in any
case, as an order of magnitude estimate of the allowed region.

We want also to mention that in the described scenario is not only possible
to seed the galactic dynamo (according to
eq.(\ref{condition})),
 but also to
produce directly fields of the same strength as that of the
galactic magnetic field, by
requiring \cite{Turner} $r(\omega_{G})\simeq 10^{-8}$. In our case
this is achieved if $z_s>10^{21}$ and if the coupling is sufficiently
tiny at the end of the dilaton driven era, $\log_{10}g_s<-43$.  The
possible impact of such a small value on the problem of freezing
out the classical oscillations of the dilaton background, in the
presence of a "realistic" supersymmetry breaking,
non-perturbative potential, is left as an interesting subject for
future research.

We want to recall, finally,  that our results were
obtained in the framework
of the tree-level, string effective lagrangian.
We know that we could  have corrections coming either from
higher loops
(expansion in $e^{\phi}$) or from higher curvature terms
($\alpha^{\prime}$
corrections ).
Since we work in a range of parameters where the dilaton is
deeply  in his perturbative
regime
 we expect our results to be stable against  loop
corrections, at least for scales leaving the horizon during the
dilaton driven phase.
As to the $\alpha^{\prime}$ corrections, they are instead invoked
in the  basic assumption that the dilaton-driven era ends when
the curvature reaches the string scale $\lambda_{s}^{-2}$, and
leads to a quasi-de Sitter epoch.
 It should be clear, nevertheless, that the basic ideas of our mechanism
and its implementation do not depend too strongly on the detailed
features of the string epoch.

\section*{Acknowledgements}

It is a pleasure to thank R. Brustein and V. Mukhanov for a fruitful
collaboration
on the spectral
properties of metric perturbations in string cosmology, which
inspired part of this work.

\section*{Note added}

While this paper was being written, we received a paper by D.
Lemoine and M. Lemoine, "Primordial magnetic fields in string
cosmology", whose content overlap with ours for what concerns
the effects of dilaton-driven inflation on the amplification of
electromagnetic perturbations. Their model of background does
not include, however, a sufficiently long, intermediate stringy
era
whose presence is instead crucial to produce the large
amplification discussed here.

\newpage

\section*{Figure captions}

\noindent
{\bf Fig. 1}

\noindent
Plot showing the evolution of the horizon scale $H^{-1}$ (thick
lines) and of the coupling scale $g=e^{\phi/2}$
(dashed lines) while the
galactic scale $\omega_G^{-1}$ (thin line) was beyond the
horizon.
 Two case with a different $z_s$
are compared, showing that, for a sufficiently fast variation
of the
dilaton during the string era, a larger $z_s$ helps achieving
the bound $g_{ex}(\omega_G) < 10^{-33}$.

\vskip 1 cm

\noindent
{\bf Fig. 2}

\noindent
The shaded area represents the allowed region determined by
the conditions (\ref{condition}) and (\ref{crit1}), and defines the
values of $z_s$, $g_s$ compatible with an amplification of the
electromagnetic vacuum fluctuations large enough to seed the
galactic magnetic field.

\end{document}